# Science Facing Interoperability as a Necessary Condition of Success and Evil

Rémy Demichelis
Université Paris Nanterre

## Abstract

Artificial intelligence (AI) systems, such as machine learning algorithms, have allowed scientists, marketers and governments to shed light on correlations that remained invisible until now. Beforehand, the dots that we had to connect in order to imagine a new knowledge were either too numerous, too sparse or not even detected. Sometimes, the information was not stored in the same data lake or format and was not able to communicate. But in creating new bridges with AI, many problems appeared such as bias reproduction, unfair inferences or mass surveillance. Our aim is to show that, on one hand, the AI's deep ethical problem lays essentially in these new connections made possible by systems interoperability. In connecting the spheres of our life, these systems undermine the notion of justice particular to each of them, because the new interactions create dominances of social goods from a sphere to another. These systems make therefore spheres permeable to one another and, in doing so, they open to progress as well as to tyranny. On another hand, however, we would like to emphasize that the act to connect what used to seem *a priori* disjoint is a necessary move of knowledge and scientific progress. *This article was presented during The Society for Philosophy and Technology Conference (June 28-30, 2021).*

## Introduction

Why are we shocked that playing video games can undermine our credit score? Or maybe you are not. But you may be shocked because of some other correlation made possible with artificial intelligence (AI), with the help of statistics, that leads to real life consequences. For example, are you okay with your insurance company having access to your smart-watch so that your insurer can check the number of steps you walk each day and, depending on your blood pressure and many other variables among which your supposedly risk of heart disease, set the price of your plan?

These scenarios may sound surprising but are actually very close to reality: "Someone who plays video games for ten hours a day (…) would be considered an idle person", said Li Yingyun, Sesam's technology director (Botsman, 2017). This Chinese program is indirectly owned by Alibaba. It has developed a social credit system that monitors people's behavior such as: one's credit history, solvency, level of education, social connections or online activity which may include political comments. A good score with this system offers an easier access to a loan and many other opportunities. It is part of a more global Chinese policy to monitor every aspect of social life, and many Social Credit Systems are currently implemented across the country though there is not a unique one. There are even possibility of sanctions with some of them that may include forbidden access to airplanes, high speed train and the impossibility to be promoted as a State official, or even your kids may not be eligible for some private schools because of your score (Arsène, 2021). It is also not clear how and if video game information is being used or if it is just a working hypothesis.

We do not necessarily have to go to China to find this kind of interoperability which leads to "surveillance capitalism". The Canadian insurance company Manulife (with Discovery) is now providing Apple Watch for free to some of its health or term life customers through the program called "Vitality". This program explicitly offers points depending on your activity that can be monitored with the Apple Watch. "Live healthy, earn rewards" says the ads. Going to the gym, taking a 15 minute-walk, getting a check-up at a doctor's office provides "Vitality points". Then, customers can obtain rewards with them: in the best case, they "can earn up to half off a hotel stay for any two nights at qualifying hotels", according to the website. But the program also offers Amazon gift cards up to 50 Canadian dollars (approx. $40 USD). Manulife Vitality does not however set the price of your plan depending on information collected through the Apple Watch, but we understand that this could be the next step, if not for Manulife, but for another insurance company. Maybe, you are still not shocked by these new possibilities. But there has to be a line AI should not cross, don't you think?

# 1. Spheres of justice

The American philosopher, Michael Walzer developed a theory according to which tyranny emerges when the goods of a certain sphere of our social life (money, religion, education, family, etc.) become predominant in another sphere. He explains in his book Spheres of Justice, in Pascal's words, that: "The nature of tyranny is to desire power over the whole world and outside its own sphere" (Walzer, 1983, p. 18).

To put it analytically (this is where we need to pay attention): sphere A may become tyrannical if a good A entitles someone to acquire a good B in another sphere B, and, more especially, if the possession of this first good A becomes mandatory to acquire the good B.

Walzer (1983, p. 10) wrote: "Every social good or set of goods constitutes, as it were, a distributive sphere within which only certain criteria and arrangements are appropriate." Money should not permit to acquire grace in the sphere of religion, nor family bonds should be a sufficient reason to obtain a diploma in the sphere of education. Spheres should have at least some autonomy toward each other.

When a sphere's rationality is denied, there is more than just tyranny, we should say. There is a feeling of injustice: the genuine experience that something went wrong because it is bad. With new technologies and the growing use of AI, we constantly risk creating such situations of tyranny or injustice. Because a machine learning system may be satisfied with some correlations to give a result that we would not accept, us, as humans. Statisticians often say "correlation is not causality", but many machine learning techniques do not need causality, unless engineers force the software to compute otherwise. Looking for causality is not what machine learning has been made for at the very beginning: it was simply asked to find patterns. Furthermore, it was set to find patterns we could not detect and therefore was freed from any causality path we knew. Nowadays, things are changing, and engineers are trying more and more to either keep the machine on some track, or at least to interpret what were the determining parameters. But in the end the job is still the same for the machine learning software: find new patterns. Because new patterns allow us to draw more precise predictions, and we always want to be more accurate.

Interoperability permits finding these patterns in more and more data, more and more spheres. It therefore creates permeability between spheres. Hence, the machine can at some point spot a criterion in one sphere that will be determining for giving a new prediction in another sphere. If we think of criterions as walzerian "goods", then a good in one sphere may entitle us to a good in another sphere according to the software no matter the moral acceptability. Because computers do not care about spheres' autonomy. Computers do not care about what goods should or should not have weight in a sphere regarding to its morally required autonomy. Spheres' autonomy is from the very beginning denied by machine learning systems. If we don't keep the software from drawing some correlations it will make them at the expense of our moral values.

We have very often been shocked by such correlations since the wide adoption of AI. Let's just think about the Amazon recruitment tool that gave men better chances to be hired because of the vocabulary they used (Dastin, 2018) or the recidivism risk assessment tool COMPAS that overestimated the risk with black people and underestimated the risk with white people (Angwin *et al.*, 2016).

But the problem goes beyond sexism and racism because an AI system may actually be very accurate in its prediction, gender and race apart. Some correlations are so strong that we cannot reject them because of inaccuracy. We may still feel some injustice even when technology is not used to reinforce domination. And this feeling of injustice is generated with spheres' permeability.

## 2. Spheres permeability

We often would rather not consider two facts together in order to keep them apart, we would rather ignore the correlation. This separation grants not only autonomy to a sphere, but also fundamental freedom to the individuals. In some occasions, even if we think there is a risk that someone does something, we would rather give the liberty to the person to act otherwise, or to do whatever she wants. The level of chance for the prediction may be very high or very low, but it does not matter. For instance, think about an early release from jail. Think about an average student that is trying hard to renew his or her scholarship. It's not just about defeating the odds. Think about excellent candidates competing for a highly selective job, some of you may know that. Some of you may know the feeling of being rejected based on non-professional criterions, like personal habits. Imagine that we tell you it is actually more accurate to take them into account. Therefore, you may have achieved more work than the selected candidate, but see she or he sticks to the "5 a day" diet. The person you are talking to may even add: studies have shown that healthy diet leads to better performance.

You may have noticed that we are not talking about AI anymore. Because any AI system is just doing what has always been asked of empirical sciences: find patterns. And this task, whether we like it or not, only requires first correlation. If at the very beginning of knowledge there is experience, there is also correlation: the genuine experience of a correlation. The skeptical philosopher from the eighteenth-century David Hume (2014) explained it clearly: "All belief of matter of fact or real existence is derived merely from some object, present to the memory or senses, and a customary conjunction between that and some other object." He admits however that we also need some causality path: we need to know which object comes first. So, he defines a cause as a certain kind of correlation: "An object, followed by another, and where all the objects similar to the first are followed by objects similar to the second." What follows logically, though maybe not expressed by Hume, is that: this causal path relies on some subjective assumption produced by imagination (Kant will talk about "plans" produced by "reason"). This assumption can nonetheless perfectly be the product of an intersubjective agreement and be a means to reproduce the observation. This is how empirical science goes on.

As the computer scientist Judea Pearl, and co-author Dana Mackenzie, wrote: "Causal questions can never be answered from data alone […] Data interpretation means hypothesizing on how things operate in the real world." But at the very beginning we are always left with correlations that do not necessarily show their chronology. We can very easily mistake an effect for a cause, but as long as we do not have further evidence to prove otherwise, this mistake is our truth. Pearl and Mackenzie admit that "after we have controlled for a sufficient deconfounding set, any remaining correlation is a genuine causal effect." They surely try to save the notion of causality, but they say at the same time that it is grounded on correlation. That means there is always room for spurious correlations.

The articulation of the laws of nature is just made of systematizing correlations. When James Clerk Maxwell, in the nineteenth century, connected electricity to magnetism, he barely (but brilliantly) put into equations some correlations. At some point there are one or few unsurpassable correlations that ground a domain's empirical knowledge. Putting them into the equations does not give them more dignity. It just makes them more predictable than other correlations. Empirical sciences progress with a better understanding, and better connections of prior knowledge.

This process is often made possible by contradicting some previous theories, but not all of them. Not the ones we still rely on, the giant-shoulder ones. When we are trying to understand a phenomenon, we are articulating these old-but-robust-theories in a manner that is not obvious. Physics has so often contradicted our most common experience by relying nonetheless and more precisely on other aspects of our knowledge that is itself grounded on the mere experience of correlations. When an AI system finds a pattern between spheres, the machine is just doing what science always did: connecting dots that were usually separated. "To understand: it is to connect knowledge, or even entire domains *a priori* separated", wrote the French physicist Hubert Krivine (2018).

## Conlcusion

We do not understand what is going on in some algorithms, but as long as they give correct predictions, some of us may be satisfied with the lack of further explanations. Because the explanation is the systematic correlation itself. Actually, the software does what Maxwell did, except that the computation is very often based on statistics and cannot be computed by hand within a human lifetime. Therefore, AI's deep ethical problem is science's deep ethical problem as well. Spheres' permeability appears to be a necessary condition of success for science. But in connecting dots between domains that were *a priori* separated, it may create situations of tyranny, or feelings of injustice.

*This article was presented during The Society for Philosophy and Technology Conference (June 28-30, 2021).*

Science Facing Interoperability## Acknowledgments

This article would not have been possible without the remarkable support of my doctoral advisors Christian Berner and Alberto Romele, my dear friend Daniel Han and, most of all, Audrey. Discussions with Tamar Sharon (whose work I highly recommend if you are interested in spheres permeability), Rua Williams and Magdalena Krysztoforska were incredibly helpful and - of course - the Society for Technology and Philosophy really made it happen.## References

Angwin, Julia, Surya Mattu, Jeff Larson and Lauren Kirchner (2016). "Machine Bias". *ProPublica*. https://www.propublica.org/article/machine-bias-risk-assessments-in-criminal-sentencing?token=-yUF56AxJcFQBoS_MWQ1wIGE6H8f7WO-

Arsène, Séverine (2021). "Le système de crédit social en Chine". *Réseaux*, 225: 55-86 : https://www.cairn.info/revue-reseaux-2021-1-page-55.htm

Botsman, Rachel (2017). "Big Data Meets Big Brother as China Moves to Rate Its Citizens". *Wired* UK. https://www.wired.co.uk/article/chinese-government-social-credit-score-privacy-invasion

Dastin, Jeffrey (2018). "Amazon Scraps Secret AI Recruiting Tool That Showed Bias against Women". *Reuters*. https://www.reuters.com/article/us-amazon-com-jobs-automation-insight-idUSKCN1MK08G

Hume, David (2014). *An Enquiry concerning human understanding*. First published in 1748. Ed. Tom L. Beauchamp. Oxford, UK: Oxford University Press.

Krivine, Hubert (2018). *Comprendre sans prévoir, prévoir sans comprendre*. Paris, France: Cassini.

Pearl, Judea and Dana Mackenzie (2018). *The Book of Why: The New Science of Cause and* Effect. Allen Lane.

Walzer, Michael (1983). *Spheres Of Justice*. New York, NY, USA: Basic Books.5